# Synchronous functional magnetic resonance eye imaging, video ophthalmoscopy, and eye surface imaging reveal the human brain and eye pulsation mechanisms

Ebrahimi Seyed-Mohsen[1,2], Tuunanen Johanna[1,2], Saarela Ville[4], Honkamo Marja[4], Huotari Niko[1,2], Raitamaa Lauri[1,2], Korhonen Vesa[1,2], Helakari Heta[1,2], Kaakinen Mika[2,3], Eklund Lauri[2,3], Kiviniemi Vesa[1,2]

[1] Oulu Functional Neuroimaging (OFNI), Department of Diagnostic Radiology, Oulu University Hospital, Oulu, Finland.
[2] Medical Imaging, Physics and Technology (MIPT), Faculty of Medicine, University of Oulu, Oulu, Finland.
[3] Faculty of Biochemistry and Molecular Medicine, University of Oulu, Oulu, Finland.
[4] Department of Ophthalmology, PEDEGO research unit and medical research center, University of Oulu and Oulu University Hospital, Oulu, Finland.

**Corresponding Author**

Vesa Kiviniemi, Prof, MD
vesa.kiviniemi@oulu.fi
Oulu Functional Neuroimaging/MIPT/ Medical Research Center (MRC)/Department of Diagnostic Radiology, Medical Research Center (MRC), Oulu University Hospital, Kajaanintie 50, 90220, Oulu, Finland.

Seyed Mohsen Ebrahimi, PhD candidate
mohsen.ebrahimi@oulu.fi
Oulu Functional Neuroimaging/MIPT/ Medical Research Center (MRC) /Department of Diagnostic Radiology, Oulu University Hospital, Kajaanintie 50, 90220, Oulu, Finland.

**Abstract**

Recent research shows the eye has a paravascular solute transport pathway driven by physiological pulsations resembling the brain. we developed synchronous multimodal imaging tools aimed at measuring the driving pulsations of the human eye. We used an eye-tracking functional eye camera (FEC) compatible with magnetic resonance imaging (MRI) for measuring eye surface pulsations. Special optics enabled the integration of the FEC with a magnetic resonance compatible video ophthalmoscopy (MRcVO) for simultaneous retinal imaging along with functional eye MRI imaging (fMREye) reflecting BOLD (blood oxygen level dependent) contrast. Upon optimizing the fMREye parameters, we thus measured the power of the physiological (vasomotor, respiratory, and cardiac) eye and brain pulsations by fast Fourier transform (FFT) power analysis. The human eye proved to pulsate in all three physiological pulse bands, most prominently in the respiratory (RESP) band. The FFT power means of physiological pulsation for two adjacent slices was significantly higher than in one-slice scans (RESP1 .vs RESP2; $df = 5$, $p = 0.0174$). FEC and MRcVO confirmed the respiratory pulsations at the eye surface and retina. we conclude that the human eye has three pulsation mechanisms, and multimodal imaging offers non-invasive monitoring of their effects in driving eye fluidics.

## Introduction

Despite the high metabolic activity of eye and brain tissues, and their consequent requirement for fluid homeostasis and metabolite clearance, lymphatic vessels are absent from the eye and brain. However, recent work shows that the eye, much like the brain, is endowed with a glymphatic drainage system that mediates metabolite and fluid clearance along perivascular spaces adjacent to the optic nerve (ON)[1]. The glymphatic brain solute transport system convects solutes along perivascular cerebrospinal fluid (CSF) spaces, being driven by physiological pulsations that predominate during sleep[2,3]. Indeed, the glymphatic system of the murine eye can remove injected tracers, as driven by intracranial pressure differences between the eye and the intracranial space[4] In human patients with normal pressure hydrocephalus (NPH), opposing MRI tracer movements have been detected around the ON after intrathecal administration of $Gd^{3+}$ contrast medium[5]. However, intra-ocular tracer injection studies are not clinically feasible for the human eye, and furthermore, the tracer molecules would likely impair the glymphatic convection by obstructing microscopic structures in the ON, especially in disorders affecting eye solute transfer and pressure[6]. Furthermore, MR (magnetic resonance) tracer studies are unfit to detect the physiological drivers affecting glymphatic convection in the human eye or brain. This technical limitation calls for developing faster, non-invasive, and tracer-free methods to image and monitor glymphatic systems[7].

Functional magnetic resonance imaging (fMRI) utilizing the T2* weighted signal offers just such a non-invasive, tracer-free approach to investigate the driving forces of CSF solute transport in the human brain. Ultrafast magnetic resonance encephalography (MREG) has recently indicated that CSF convection in human brain is affected by three drivers, namely vasomotor, respiratory, and cardiac[8] pulsatility, each of which increases in power during slow wave sleep in brain regions of increased CSF convection [9]. Importantly, the physiological pulsations are differentially involved in divers neuropathologies, including Alzheimer's disease[10], epilepsy[11], primary central nervous system (CNS) lymphoma[12] and narcolepsy[13]. However, the ultrafast MREG (magnetic resonance encephalography) spiral acquisition scans established for brain are not optimized for eye tissue characterization.

Echo planar imaging (EPI) of blood oxygen level dependent (BOLD) contrasted fMRI has been used to correlate brain function to eye gaze, pupil dilation, and eyelid closure as markers of numerous brain functions including attention, sympathetic activity, and vigilance[14]. The functional

BOLD fMRI of eye tissue (fMREye) technique has hitherto focused on effects of voluntary eye movement and eyelid opening and closure[15]. Duong and colleagues have pioneered a retinal BOLD imaging method that reveals classic hyperemic BOLD responses in the retina during visual stimuli and physiological challenges[16]. EPI in one or two neighbouring slices can be imaged with high in-plane resolution to enable the delineation of small intraocular structures, also with sufficiently high temporal resolution (100 ms) to sample physiological pulsations without need for aliasing[17]. Despite its non-invasive character and the wide availability of fMRI, there are no reports regarding the physiology of human eye pulsatility as revealed with BOLD contrast imaging.

In this work, we developed and optimized a gradient echo EPI for fast fMREye with a scanning sequence of 100 ms repetition time to detect eye and ON pulsations. To confirm and quantify the physiological eye pulsations, in addition to conventional peripheral measures of cardiorespiratory physiology, we developed two MRI-compatible procedures to measuring simultaneously and more directly the physiological pulsations of the eye during MRI scanning. First, we used a functional eye camera (FEC), which is commonly employed for eye/pupil tracking in fMRI studies, to detect corneal surface pulsatility in synchrony with the fMREye measurements[18]. Second, we further augmented the FEC with advanced optics to provide a dedicated MRI-compatible video ophthalmoscope (MRcVO), which we used to scan retinal pulsations during fMREye imaging[19]. In our optimization of the fMREye imaging parameters, we aimed for minimal interference with the MRcVO and FEC camera signals. All three physiological pulsation frequency bands in fMREye, MRcVO, and FEC signals were verified against conventional physiological monitoring data of the subjects. With the introduction of this technology, we show pioneering findings regarding physiological pulsations that drive solute transport in the human eye.

## Results

### Number of slices

First, we optimized the number of eye slices, aiming to minimize signal artifacts and to maximize signal power. Next, we investigated whether fMREye could also detect all three physiological pulsation types from intracranial space. From the whole head image data, the FFT power spectra could synchronously identify the three distinct eye and brain pulsation bands (Fig. 1B-D), namely the VLF, respiratory and cardiovascular frequencies as known from earlier studies. To optimize the number of slices within the ultrafast 100 ms TR scanning timeframe, we obtained recordings in one and two slices. Subjects were scanned after pupillary dilation of the left eye in MRcVO (n ± 12) and without dilation in FEC (n = 6). The two slice data sets had significantly higher mean BOLD signal intensity level and pulsation power (Fig. 1). For each of the three pulsations bands, the peak of FFT power for 2 slices exceeded that of single slice recordings. The fMREye scan robustly detected all three physiological brain pulsations, i.e., vasomotor, respiratory, and cardiovascular pulsations, c.f. Fig. 1E.

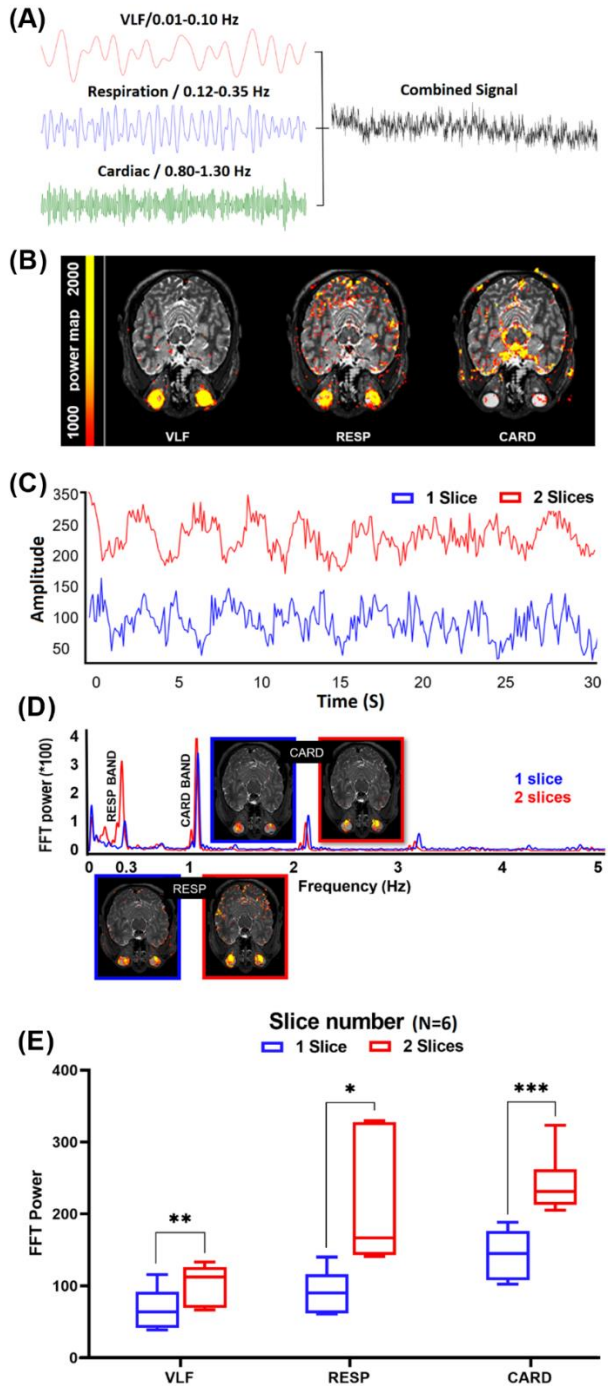

***Figure 1.*** *A: A single-subject map example of raw fMREye signal with filtered very low frequency (VLF 0.01-0.1 Hz) and individual bands for respiration (RESP 0.12 – 0.35 Hz) and cardiovascular (CARD 0.8 – 1.3 Hz), which were 0.1 Hz wide and centered around the individual peaks (i.e., peak ± 0.05 Hz). B: An example of the individual FFT power map of the three physiological pulsations VLF, RESP, CARD). C: fMREye time signal example from one voxel of the retina during one and two slice fMREye scanning: The two-slice signal was of higher amplitude than the one slice scan. D: The mean FFT power plot for the whole head fMREye image showing power peaks in three physiological bands. E: Statistical chart for data in six subjects showing that the sum of FFT power*

*for two slices (VLF$_2$, RESP$_2$, CARD$_2$) exceeded that for one slice scans in all three physiological pulsation bands (VLF$_2$, RESP$_2$, and CARD$_2$ versus (VLF$_1$, RESP$_1$, CARD$_1$).*

The mean FFT power (SD) in arbitrary units for one slice VLF was 68.11 ± 28.73 (VLF$_1$) versus 103.1 ± 28.24 (df = 5, p = 0.0087**) for VLF$_2$.; Corresponding values for RESP were 91.89 ± 30.21 versus 212.4 ± 90.73 (df = 5, p = 0.0174*); and for CARD were 143.6 ± 34.70 versus 241.4 ± 42.57 (df = 5, p = 0.0007***) (Fig. 1E).

**MR-compatible MRcVO and FEC camera signal interreference**

We next investigated the effect of different slice numbers on the MR-compatible eye camera signals. Single slice fMREye scanning had a large artifact effect in the FEC/MRcVO camera signal compared to two slices fMREye data, as shown in Fig. 2.

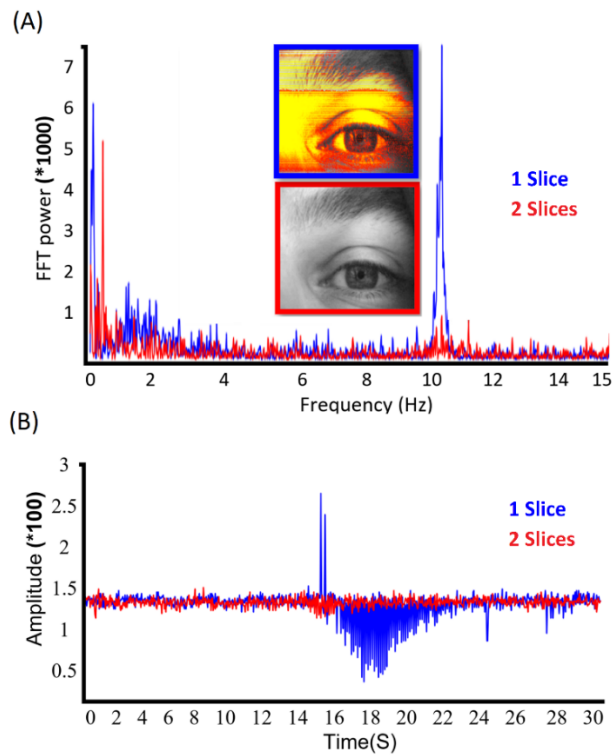

*Figure 2. Lower image flicker artifact in two slices mode on eye images obtained by FEC during the fMREye scan. A) FFT power chart shows in the 10 Hz frequency band a high noise power in the one slice scan FFT spectrum, which is almost absent in the corresponding two slice FFT spectrum. B) A 30 sec time signal for the eye corneal surface FEC video data, illustrating artifactual flicker for the single slice (blue) (fMREye signal compared to the more stable two slice fMREye scan (red).*

The one slice signal data had markedly more flicker in the video recording (Fig. 2B). The peak amplitude for the flicker artifact in the FFT power spectrum was 10 Hz. However, this artefact was clearly absent in the 2-slice data, which we therefore selected for further analysis.

**Slice thickness optimization**

To optimize the slice thickness of slice, we scanned the fMREye data for three difference slice thicknesses: 3, 4 and 5 mm. Obviously, thinner slices have better spatial resolution because of smaller voxel size and lesser effects of partial volume. Also, MRI resolution for two slices of 3 mm slice thickness would certainly be better for one slice of 6 mm slice thickness, but with comparable coverage of the ON.

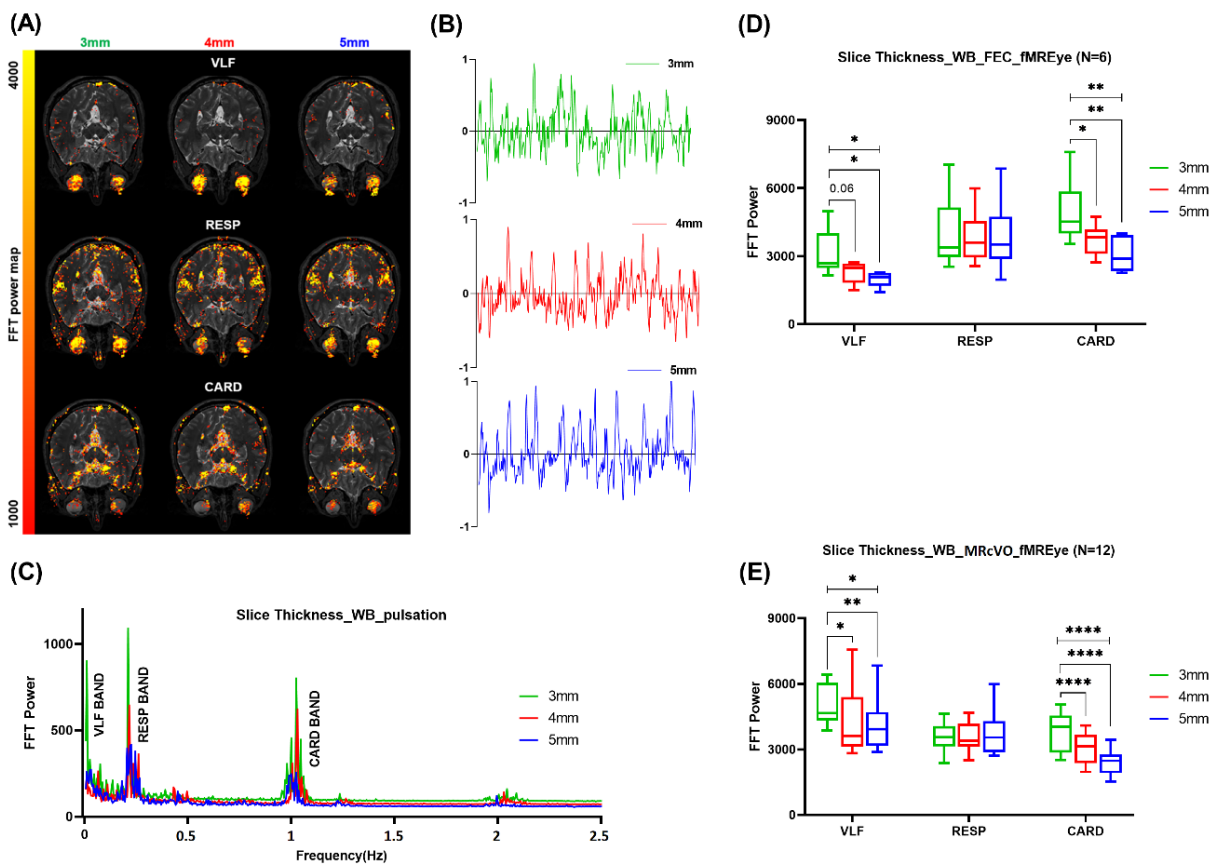

*Figure 3.* fMREye scan detects vasomotor (0.01-0.1 Hz), respiratory (0.25 ± 0.05 Hz), and cardiovascular (1.1 ± 0.05 Hz) pulsations both from whole brain and eye structures. A: A representative FFT power map of the physiological pulsation from two slice data of 3-, 4-, and 5- mm slice thicknesses. B: The fMREye time signal from one voxel of the retina for different slice

*thicknesses. C: The mean FFT power of the same subject's whole fMREye image for all scan protocols illustrate clear VLF, RESP, and CARD peaks and cardiorespiratory harmonics. D: The FFT power spectrum in different slice thickness from whole brain (WB) fMREye simultaneous FEC, without pupillary dilation. The mean FFT power in the cardiac band was significantly higher for thinner slices, and likewise in VLF band for 3 mm slices compared to 5 mm. There were no significant differences in the RESP band. E: The FFT power spectra from whole image fMREye data of different slice thickness, obtained simultaneously with MRcVO after pupillary dilation. FFT power for VLF and cardiac pulsations were higher for thinner slice scans compared to 5 mm thickness. ($* > 0.05$, $** > 0.01$, $*** > 0.001$, $**** > 0.0001$).*

Next, we investigated the effects of slice thickness on FFT power and signal amplitude. Thinner slices proved to have stronger signal power and FFT power amplitude. We found significant differences for CARD and VLF results in both conditions (whole eye and brain scanned simultaneously by fMREye with eye scanning by FEC and retina scanning by MRcVO), but no significant results for RESP according to the ANOVA statistical test results. Inspection of data for the whole image shows conspicuously higher power in the VLF and cardiac bands for 3 mm slice thickness compared to 4- or 5-mm thickness. In this population, we saw no effect of slice thickness on the respiratory frequency-related pulsation FFT power.

Mean (SD) FFT power values (n = 6 subjects, df = 5) in whole image FEC and fMREye data for VLF was $3128 \pm 1043$ at 3 mm slice thickness ($VLF_3$), $2299 \pm 479.0$ at 4 mm slice thickness ($VLF_4$), and $1986 \pm 330.1$ at 5 mm slice thickness ($VLF5_5$) (ANOVA test, p = 0.046). $VLF_3$ vs. $VLF_4$ (Student's *t*-test, p = 0.061), $VLF_3$ vs. $VLF_5$ (Student's *t*-test, p = 0.047). Corresponding mean FFT powers were $3998 \pm 1627$ for $RESP_3$, $3814 \pm 1217$=for $RESP_4$, and $3842 \pm 1646$ for $RESP_5$ (ANOVA test, p = 0.72). Corresponding mean FFT powers were $49,229 \pm 1435$ for $CARD_3$, $3723 \pm 687.9$ for $CARD_4$, and $3051 \pm 750.0$ for $CARD_5$ = (ANOVA test, p = 0.0048). $CARD_3$ vs. $CARD_4$ (Student's *t*-test, p = 0.019), $CARD_3$ vs. $CARD_5$ (Student's *t*-test, p = 0.0086), Fig. 3d.

In the whole image MRcVO and fMREye data, the mean (SD) FFT power (n =12 subjects, df = 11) was $5079 \pm 878.6$ for $VLF_3$, $4303 \pm 1500$ for $VLF_4$, and $4161 \pm 1206$ for $VLF_5$ (ANOVA test, p = 0.016). $VLF_3$ vs. $VLF_4$ (Student's *t*-test, p = 0.046), $VLF_3$ vs. $VLF_5$ (Student's *t*-test, p = 0.005). Corresponding mean (SD) of FFT powers were $3509 \pm 666.2$ for $RESP_3$, $3538 \pm 660.9$ for $RESP_4$ and $3833 \pm 1095$ for $RESP_5$ (ANOVA test, p = 0.38). Corresponding mean (SD) of FFT powers were $3769 \pm 865.6$ for $CARD_3$, $3070 \pm 728.2$ for $CARD_4$, and $2410 \pm 558.4$ for $CARD_5$ = (ANOVA test, p < 0.0001). $CARD_3$ vs. $CARD_4$ (Student's *t*-test, p < 0.0001), $CARD_3$ vs. $CARD_5$ (Student's *t*-test, p < 0.0001), Fig. 3e.

**Optic Nerve (ON)**

The ON is an important channel for glymphatic eye clearance[20]. Therefore, we analyzed specifically the ON in fMREye scan data. Fig. 4 depicts a representative ON region of interest (ROI) projected on the structural MR image.

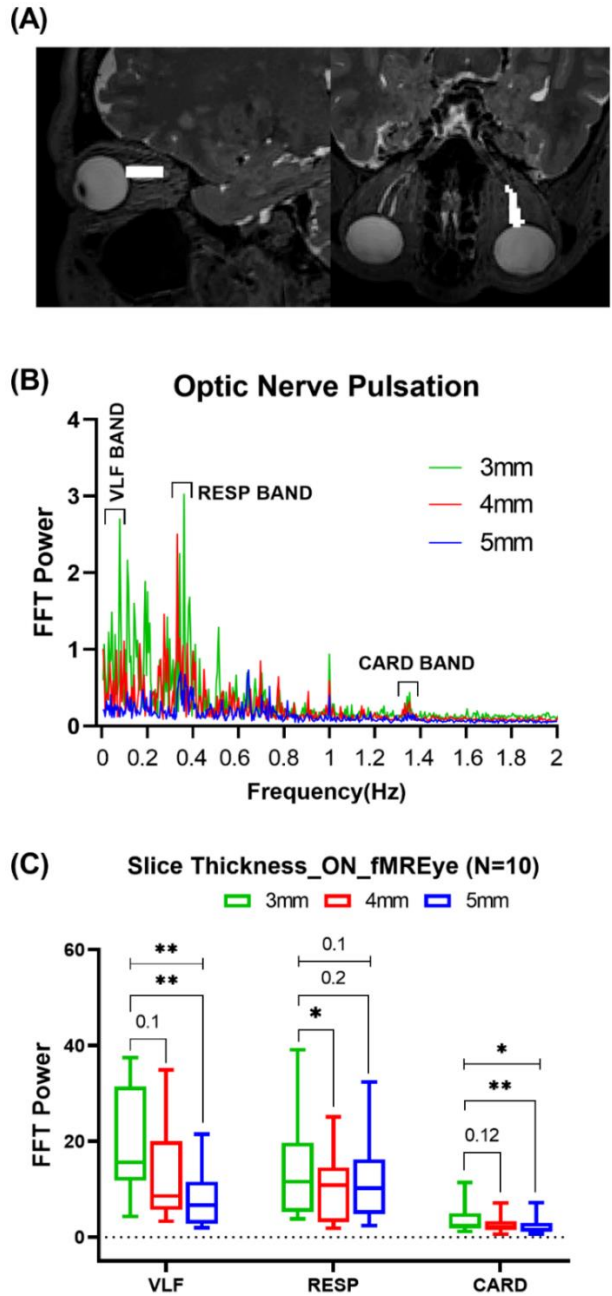

*Figure 4. A) ROI map of the optic nerve (ON); B) FFT power mean chart of the ON, showing higher power in the respiration band for 3 mm slice thickness compared to 4 and 5 mm; C) Group*

*results of the FFT power for ON pulsations, showing significantly higher power in VLF, RESP, and CARD bands for 3 mm thickness compared to 4 and 5 mm.*

The slices thicker that 3 mm usually extended beyond the ON ROI, thus being vulnerable to partial volume effects from surrounding tissue. The three pulsation bands from the fMREye data indicated significantly higher power in the VLF and CARD bands for slices of 3 mm thickness. The respiratory pulsations tended to have the lesser dependence in slice thickness (as seen for the retina, above), without reaching significance. Ten among the twelve subjects (due to lacking ON coverage for two subjects) served for investigation of the ON pulsations. One-way ANOVA showed no significant effect of slice thickness for RESP results. Mean (SD) FFT power (N = 10 subjects, df = 9) were 19.67 ± 11.36 for $VLF_3$, 12.87 ± 10.34 for $VLF_4$, and 7.73 ± 6.24 for $VLF_5$ (ANOVA test, p = 0.008), with $VLF_3$ vs. $VLF_4$ (Student's *t*-test, p = 0.102) and $VLF_3$ vs. $VLF_5$ (Student's *t*-test, p = 0.0013). Mean (SD) of FFT power was 14.44 ± 10.85 for $RESP_3$, 10.45 ± 7.47 for $RESP_4$, and 11.81 ± 9.24 for $RESP_5$ (ANOVA test, p = 0.103). Mean (SD) of FFT power was 3.74 ± 3.14 for $CARD_3$, 2.73 ± 2.02 for $CARD_4$, and 2.35 ± 2.03 for $CARD_5$ (ANOVA test, p = 0.037). $CARD_3$ vs. $CARD_4$ (Student's *t*-test, p = 0.124), $CARD_3$ vs. $CARD_5$ (Student's *t*-test, p = 0.0047), Fig. 4c.

**Human video ophthalmoscopy and fMREye BOLD signal**

To verify the existence of physiological eye pulsations, we interrogate the MRcVO and FEC scans. As expected, we indeed found synchronous eye and brain pulsations in the *respiratory band* for three individuals extending from the ocular surface to the back of the eye, as shown by fMREye and FEC, and by MRcVO imaging. The MRcVO and FEC results both confirmed that the eye does indeed pulsate synchronously with the fMREye pulsations, c.f. Fig. 5. Only 3 of 12 cases were successful for MRcVO, whereas 3 of 6 cases were successful for FEC in detecting the respiratory pulse. Some datasets were excluded due to eye and head motion, and some were usable due to other artifacts and camera interference flickers.

However, due to head motion and eye saccades, the MRcVO data were generally unsuitable for physiological FFT power analysis. *Mcflirt* motion correction of nifti-files of MRcVO data failed to align the dynamic saccade results to a satisfactory extent (c.f. supplementary MRcVO video). FEC data suffered less from eye saccades and were therefore able to capture pulsations. However, the VLF and CARD pulsation bands were not detectable, due to a) harmonics of RESP and VLF masking the CARD band and b) the existence of several overlapping pulsation peaks in the VLF band, thus impeding its accurate recognition.

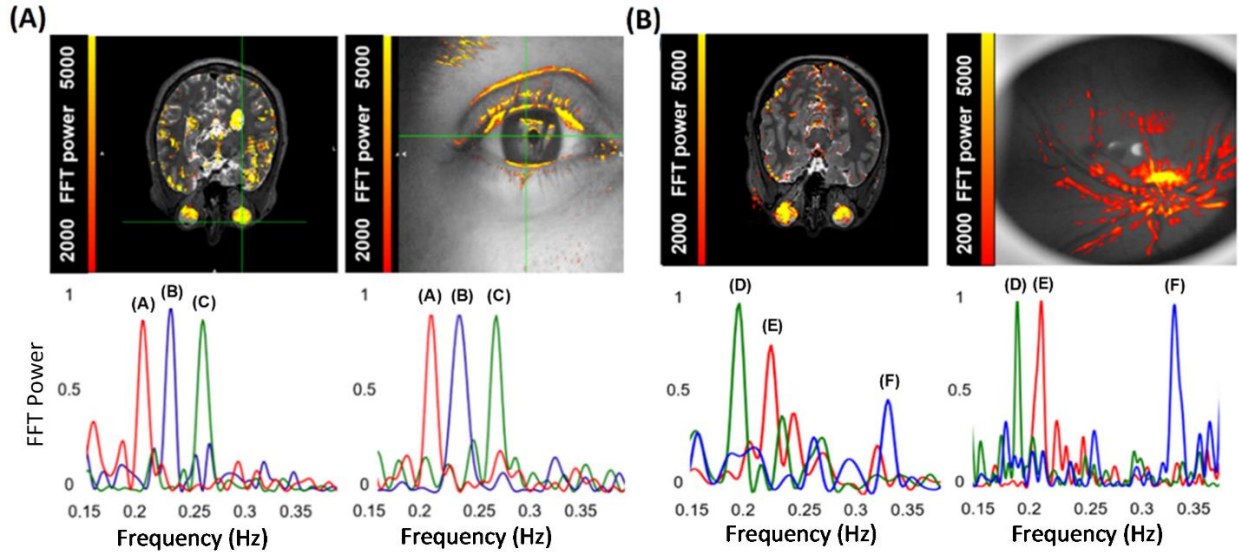

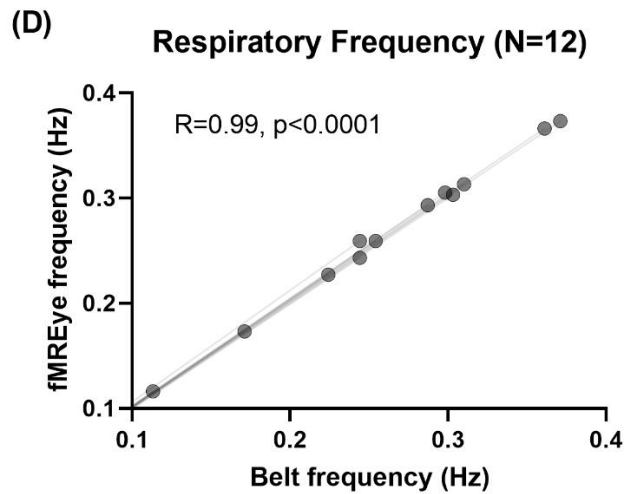
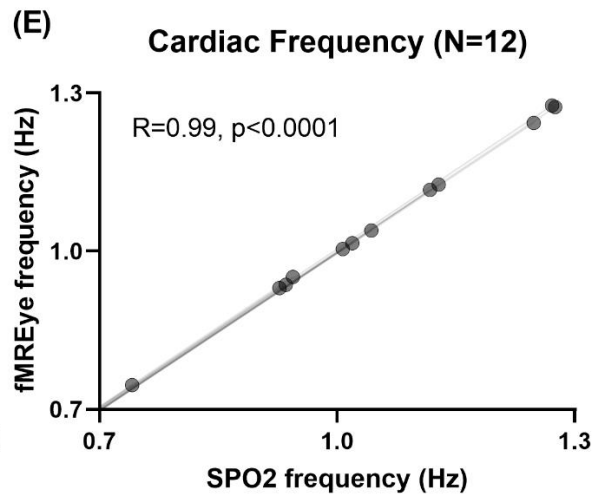

*Figure 5.* A) Simultaneous fMREye and FEC surface video scanning reveals strong respiratory eye pulsations at identical frequencies in each of three subjects. The whole eye, as well as the pupil, pulsates in the respiratory frequency. B) Simultaneous fMREye and MRcVO also detect strong respiratory range pulsations in the human eye. Notably, both the whole eye structure as well as the retinal vasculature pulsates strongly at identical respiratory frequency ranges in all three subjects. C) After checking the fMREye results with belt and oximeter recordings, a comparison of FEC and MRcVO data to fMREye imaging data showed identical peaks in the RESP band. D) Respiratory belt data and E) cardiovascular SpO2 showed identical physiological frequency peaks compared to the fMREye whole image FFT spectra.

Fig. 5 presents the results of three cases in comparing eye surface imaging with fMREye scanning, and their use for comparing the RESP band of eye pulsation in the back of the eye by MRcVO and fMREye. The peak frequencies in fMREye and FEC/MRcVO recordings matched each other (supplementary table 2). The respiratory and cardiac pulsations results achieved from fMREye were confirmed by respiratory belt data and cardiovascular SpO2 for twelve subjects (Fig. 5D and E), illustrating identical physiological frequency peaks compared to fMREye whole image FFT spectra. Statistical analysis for both graphs showed $p<0.0001$ and Spearman's correlation (R) equal to 0.99.

## Discussion

In this study, we investigated new hybrid methods for measuring physiological pulsations of the eye using functional eye MRI BOLD imaging (fMREye), simultaneously with a functional eye camera (FEC) and MR-compatible video ophthalmoscope (MRcVO). Our optimization of the scanning procedures showed that 2-slice fMREye imaging with thin (3 mm) slices presented the least artifacts and was most sensitive in detecting pulsation powers in the eye. We further verified that the fMREye pulsation data precisely reflected physiological cardiorespiratory recordings both from the eye and brain, as further substantiated by peripheral measurements. Compared to the brain pulsation data, the ON has strong VLF and respiratory pulsations, but surprisingly small cardiovascular pulsation. This result was also verified by the simultaneous FEC and MRcVO data, wherein neither the eye surface nor retina presented strong respiratory band pulsations, nor a significant cardiac pulsation in the human eye.

The successful imaging of physiological drivers of glymphatic solute transport in the eye called for specific technical improvements due to the special characteristics of the eye with respect to MRI. Advances in MRI technology and image processing such as image signal despiking enable imaging of the human eye with high spatiotemporal resolution[21]. In this study we optimized 3T conventional gradient recalled echo planar imaging parameters to measure human eye physiological pulsations. To the best of our knowledge, this is the first study to measure systematically human eye pulsations.

We detected all three of the expected pulsation frequencies in the human eye. Previous fMRI studies of the eye aimed to assess the brain effects of what might be called general eye activity such as eyelid closure[22] and saccadic movements, both of which proved to modify brain neuronal network connectivity[23] and the amplitude of resting state fluctuations[24]. The eyelid posture reflects vigilance state, which has marked global effects on brain function, and is related to the emergence of widespread co-activation patterns after sleep onset[14]. Consensual changes in the eye's pupil size are strongly associated with neuromodulator tone. More precisely, pupillary diameter is controlled by sympathetic and parasympathetic systems, and it is recognized as a peripheral index of arousal in addition to its well-known regulation by ambient illumination[25]. The pupil dilation reflects the balance of effects arising from multiple brain regions, such as the hypothalamus and locus coeruleus.[26]

Several rhythmic or diurnal processes affect the eye. Like the diurnal volume variations of the human brain, which is greater in the morning[27], the choroid of the eye is thickest at night, and its axial length shortest[28]. This is in line with the findings of increased brain glymphatic function [29] and ocular CSF clearance[1] during the night, which is mediated by aquaporin-4 water channels. It is also possible that the nocturnal increase in retinal thickness may be related to increased convection of glymphatic solutes due to increased water content.

While the brain pulsates strongly in VLF and respiratory frequency bands[8,9,30], there is limited information on corresponding eye pulsations. However, the eye is known to have a spontaneous intraocular pressure (IOP) and venous pulsations throughout the bulbus, which track the cardiac rhythm[31]. Furthermore, eye pulsations cease when the CSF pulse pressure equals the IOP [32]. We detected this cardiac pulsation in the ON with fMREye scanning, but surprisingly its power was relatively insignificant compared to that of the other two pulsations (Figs. 1,3-5). The cardiac pulse was likewise exceedingly small in both FEC and MRcVO signals compared to respiratory pulsation power.

The human ON can range in diameter from 0.96 to 2.91 mm, with atypical size of 1.5 mm [33]. To minimize partial volume effects, we require a correspondingly small slice thickness to accurately measure ON pulsations in fMREye. A two-slice scan yielded a stronger signal and higher FFT power for all three physiological pulsations compared to a one-slice scan. Also, the flickers were of significantly lower amplitude with simultaneously recorded 2-slice FEC and MRcVO signals, due to the camera interference effect arising during two-slice scanning.

After settling upon two-slice scanning, we then proceeded to compare results with 3-, 4-, and 5-mm slice thicknesses. We obtained stronger whole brain means of FFT power amplitude for VLF and CARD pulsation bands in two slices of 3 mm thickness. It is noteworthy that the RESP and CARD pulsations of the fMREye data both matched the physiological measurements of respiratory belt and cardiovascular SpO2-pulseoximeter, with perfect agreement ($R = 0.99$ and $p < 0.0001$). A ROI analysis of the ON indicated stronger FFT power of the VLF and CARD bands for 2-slices of 3 mm thickness. However, respiratory power in the ON was not affected by the slice thickness in any of our measurements.

The orientation of the ocular axis, i.e., the gaze direction, within the scanner is especially important during the fMREye recordings since the orientation of the ON will follow the ocular axis[34]. A neutral position of the eye axis would allow the slice positioning to have the widest coverage of the ON, importantly covering its entire diameter including the sheath. When the participants are looking up or down, the coverage of the ON declines, such that pulsation results may change due to transient stretching of the ON and ocular structures[35].

Eye tracking by EEG[36] or a MR-compatible video camera[37] are commonly employed to monitor pupil diameter, gaze direction, or vigilance during MRI[38]. MR-compatible eye tracking cameras record gaze position during scanning with high temporal and spatial resolution, and hence enable the investigation of gaze-related brain activity. In practice, however, camera systems are seldom used in fMRI studies[39], often due simply to their unavailability in research or clinical settings. Indeed, MR-compatible cameras are expensive, require trained staff, and costly setup and calibration time, while imposing experimental constraints, such as the need for eyes open[40]. However, we consider this technology to augment importantly the scope of fMRI investigations.

We now used FEC to detect physiological eye pulsations during the fMRI brain acquisition, following upon pioneering studies of Li et al. employing a frontal eye tracker to record pulsatility of the eye surface[41] and likewise the pupil edge. We find that FEC captures the respiratory pulse on the orbital surface, but the cardiac pulse is elusive due to interference from respiratory pulse harmonics in multiples of the principal frequency of respiration. We were able to filter out some undesirable reflections from the eye's surface, eyelid, and face skin surface using the *fslroi* function. In three subjects FEC and fMREye simultaneously detected the same strong respiratory range pulsations in the human eye.

Previously, retinal video sequences have been proposed to be useful for the observation of spontaneous venous pulsation (SVP), a phenomenon arising from changes in the pressure gradient between intraocular and intracranial pressure during the cardiac cycle. SVP is usually observable with the ON head in healthy eyes[42]. In this study, we analyzed the capability of simultaneous fMREye and MRcVO for detecting physiological eye pulsation mechanisms. Based on classical understanding of neuro-ophthalmic pulsations, we predicted that cardiac pulsation should be the dominant feature in the eye. However, we were unable to confirm this hypothesis with MRcVO so far, possibly due to masking of the pulsation by eye movements, given that such pulsations are indeed observable well-controlled video ophthalmoscopy outside the Faraday cage [43].

Motion artifacts from eye saccades present a big challenge for recording eye pulsatility by MRcVO. We applied FSL *Mcflirt* to all imaging data (fMREye, FEC, and MRcVO) after transformation into nifti format, but this approach was unfit to correct the saccade motion in most FEC/MRcVO recordings. This may be due to the optimization of *Mcflirt* for 3D BOLD image registration rather than the present 2D imaging. We used AFNI *3dDespike* for removing effects of eye lid closure during blinking, which robustly removed the image spikes related to eye lid blinking. High-level feature descriptors such as scale-invariant feature transformation (SIFT)[44] and speeded-up robust features (SURF)[45] that are generally used retinal image registration can also serve for motion correction in pre-processing[46].

Eye movements introduced large artifacts to eye dynamic scanning, thus substantially hindering our pulsatility data analysis. The fMRI of the human retina is disfavored due to eye motion and limited spatial resolution due to weaker magnetic field gradients[47]. Plöchl and colleagues reviewed the properties of eye movement artifacts, including cornea-retinal dipole changes, saccadic spike potentials, and eyelid artifacts, and reported on their interrelations[48]. In concordance with earlier studies, they found that these artifacts arose from independent sources, with effects on the measured signal that depended on electrode site, gaze direction, and choice of reference sources.

We aim to reduce eye motion artefacts by stabilizing the ophthalmic imaging system in real-time[49]. Irrespective of various adjunct devices, using a filtering algorithm to enhance the signal-to-noise ratio increased the FFT power of the physiological pulsation bands of interest and removed eye movement artifacts[50], which presents another focus for future work[51].

By determining the correspondence of respiratory pulses in all recording devices, it should prove possible to perform FEC and MRcVO examinations outside the MR room with a general video recording to investigate eye pulsation. Also, comparing glaucoma and IIH patients with control groups presents an attractive research topic. To enable eye closed scanning and motion correction algorithms within MRcVO calls for improved motion control for the MRcVO recordings. The aquaporin-4 channels that drive glymphatic flow are affected by specific antibodies in neuromyelitis optical-spectrum diseases including bilateral optic neuritis and spinal cord myelitis, presenting another target for the present methods. Furthermore, we plan to investigate the effects of assuming a supine position on equalization of differential IOP/ICP.

There are five diverse types of eye movements contributing to motion artefacts: saccades, smooth pursuit, and vestibular, optokinetic, and vergence eye movements[52], which can produce rotations of 40° in only 100 ms[52]. How these movements affect the dynamics of ocular fluids is only vaguely understood, although micro saccadic oscillations are related to dynamic changes in the vitreous humor[53]. Eyelid closure[22] and saccadic movements also modify also downstream brain neuronal network connectivity[23] and the amplitude of resting-state brain fluctuations[24], thus presenting additional research targets in addition to the effects of eye lid motion on solute transport of the eye itself.

Commonly, efforts to reduce eye movement signals arise in the context of BOLD contrast scan signals, but these same movements may be drivers for solute movement. Thus, we perceive an opportunity to establish better the nature of eye physiological pulsation bands in a multimodal approach. Better motion control may help in obtaining more reliable data from the retina through a projection-artifact removal algorithm[48,50].

## Conclusions

We developed a simultaneous multimodal brain and eye scanning system that enables the detection of physiological neuro-ophthalmic pulsations driving glymphatic solute transport in the healthy human eye. Surprisingly, on the macroscopical scale, our several methods concurred in showing that the human eye shows stronger pulsation in vasomotor and respiratory bands than in cardiovascular frequencies. Results provide the basis for future investigations of ocular diseases such as glaucoma or brain disease such as IIH. We foresee future applications of our methods in the non-invasive measurement of intracerebral pressure, which may prove to be of pivotal medical importance.

## Materials and Methods

The study population comprised 16 healthy volunteers (5 males, 11 females, of mean age (SD (Standard Deviation)) of 38.7 ± 12.4 years). We completed the analysis in all sixteen subjects, with duplicate scans in two individuals. The study and the applied fMREye, MRcVO, and FEC technology were approved by the Regional Ethics Committee of the Northern Ostrobothnia Hospital District (FIMEA/2021/002535) The study was performed in Oulu University Hospital, University of Oulu, Oulu, Finland. We obtained written informed consent from all participants, according to the requirements of the Declaration of Helsinki. All procedures were performed in accordance with relevant guidelines. All subjects were healthy and met the following inclusion criteria for multimodal scanning: no continuous medication, no neurological nor cardio-respiratory diseases, non-smokers, and no pregnancy.

### Data acquisition

Subjects were imaged in Siemens MAGNETOM 3T Vida scanner (Siemens Healthineers AG, Erlangen, Germany) using 64-channel head-coil. We used the following scanning parameters for the Siemens ep2d GRE-EPI (2D echo planar imaging) BOLD sequence: repetition time (TR = 100 ms), echo time (TE = 15 ms), echo train length (ETL:32), flip angle (FA = 15°), 128*128 matrix yielding 2.64x2.64 mm pixels, and one or two slices of varied thickness: (3, 4, and 5 mm). For structural alignment, we obtained anatomical 3D 0.9 mm cubic voxel images from T1-MPRAGE (TR = 1900 ms, TE = 2.49 ms, FA = 9°) and T2-spc (TR = 3200 ms, TE = 412 ms, FA = 90°) with 240 mm$^3$ field of view (FOV).

Twelve of the 16 subjects (5 males, 7 females, mean (SD) age 42.5 ± 14.4 years) were scanned for 3 min epoques with simultaneous eye-scanning by either MRcVO or fMREye. The MRcVO subjects' left eye pupil was dilated with tropicamide (OFTAN TROPICAMID, 5 mg/ml) eye drops. The remaining six subjects (all males, of mean (SD) age 33.8 ± 9.1 years) were scanned with the functional eye camera (FEC) during fMREye imaging of the eye surface structures, but without pupil dilation. Two subjects underwent both MRcVO and FEC with fMREye scanning on separate days. Eye scanning commenced after the pupil had been dilated some 3-5 min after drug administration. Multimodal imaging was also used to measure the physiological pulsations, along with scanner respiratory belt and right index fingertip SpO2 pulse oximeter[54] recordings from the group of 12 subjects.

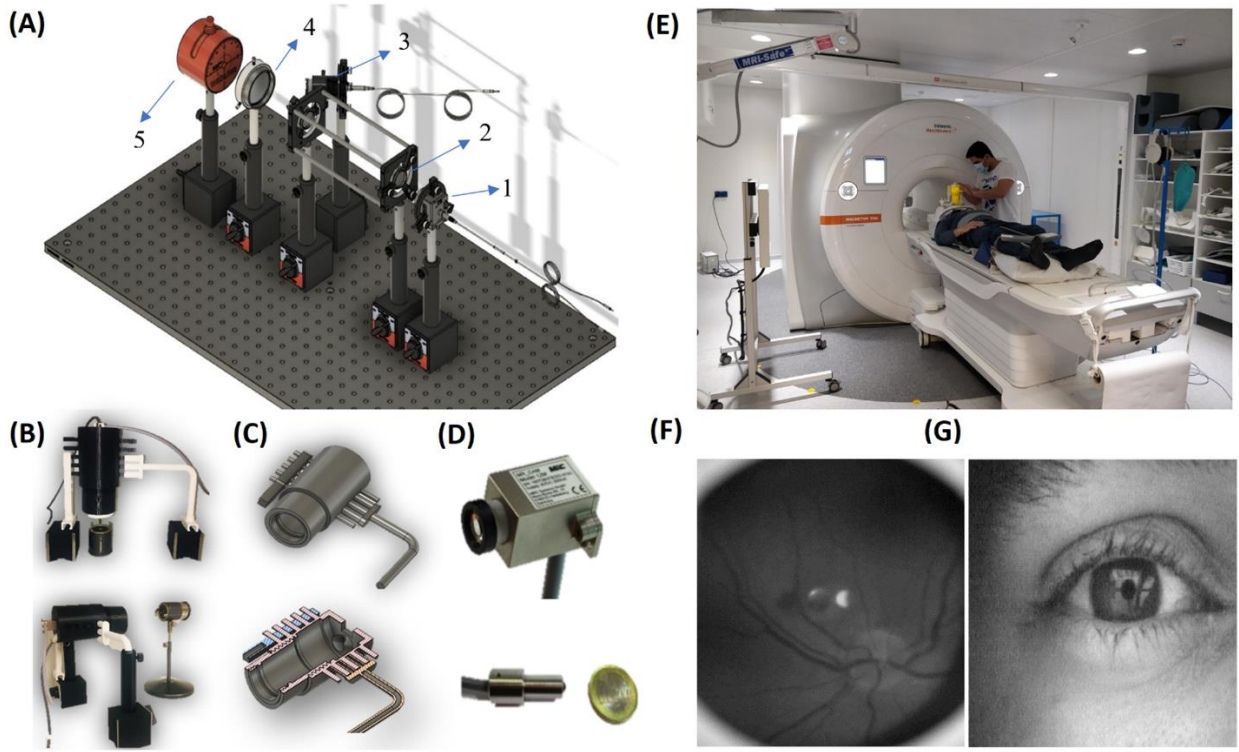

*Figure 6.* Technology and development of the eye scanning setup. A: Optical equipment for the experimental testing of fundoscopy. B: Third version of the prototype fundoscopy system for vertical and horizontal positioning of MR-compatible video ophthalmoscope (MRcVO). C: ABS material 3D-printed optical system holder for use inside the MR scanner room's Faraday cage. D: MRC's (R) MR compatible camera, and MR compatible LED (light emitting diode) light source. E: Human fMREye scanning in a 3T Vida MRI laboratory. Example of image of F) MRcVO video image illustrating the human retina, c.f. Supplementary file for simultaneous fMREye/ophthalmoscopy mpeg-video, and G) functional eye camera (FEC) video taken during simultaneous 3T fMREye scanning.

This system is the first MR-compatible video ophthalmoscope, which enables continuous video recording from the retina. The MRcVO is an optical system that includes, as shown in Fig. 6A: 1. The MR compatible video camera (MRC Systems GmbH, Heidelberg, Germany), 2. Zoom lenses 3. MR compatible white LED, and 4. a 20D/28D lens (Volk Optical Inc., Mentor, OH, USA), which were assembled inside an 3D printed acrylonitrile butadiene styrene (ABS) optical system holder of in-house construction (FABLAB, University of Oulu, Oulu), (Fig. 6B-D). Due to individual differences in pupillary dilatation, we scanned the retina with two different optical arrangements. In the first arrangement, for those with pupil diameter greater than 4 mm open, we used a 20 diopter Volk lens placed 13 cm in front of an MRC camera equipped with an 8 mm zoom lens. In this setup, the FOV was 30 degrees, with an optimal distance between the eye and the fundoscopy of about 5 cm. In the second arrangement for individuals with pupil measuring less than 4 mm, we used a 28 diopter Volk lens positioned 15 cm in front of the MRC camera equipped with a 12 mm zoom lens. For this optical setup, the FOV was 60 degrees, with the optimal distance between the eye and the fundoscopy of about 3 cm.

The light source, consisting of a white LED and a blue high-pass filter to remove wavelengths less than 500 nm, was aligned under the camera at an angle of 30 degrees, such that the reflected rays of the lens surface scattered to the surrounding side area and did not return to the camera, thus uniformly illuminating the surface of the retina with white light. In our setup, the incident light enters the eyeball from the sides of the pupil (towards the posterior part of the eye). The reflected light exits from the center of the pupil and is directed to the MRC camera by the lenses, with passage through a filter box. The filter box prevents the transmission of extraneous signals into the MR cabinet, thus minimizing interference with the video signals and MRI imaging. Signals are recorded digitally via a video card. Unlike in conventional ophthalmoscopes, we simplified our system to enable its use inside the MR bore to capture images from the retina and eye surface simultaneous fMREye (Fig. 6E-G), by placing the light source in the same path as the optical imaging system. So designed, the system occupies a smaller space, which does not interfere with installation of the Siemens 64-channel head coil.

We tested the system using an artificial eye model[55] as illustrated in Fig. 6(A). A 3D model can be found on our webpage (https://www.oulu.fi/en/research-groups/oulu-functional-neuroimaging-ofni). The continuous measurements call for low light intensity during the prolonged illumination, especially given the pharmacological pupil dilation. The MRcVO system contains a source for visible light (400 to 700 nm), as shown in Fig. 7. The spectrum of the MR-camera light source raises the possibility of blue-light photochemical and thermal injury of the eye. We obviated this risk through long-pass filtration of blue and ultraviolet light (wavelength < 500 nm) projected by the LED[56] and scrupulous control of light intensity. To meet and exceed eye safety regulations, we set the light intensity to a maximum of 2700 lux, which we confirmed by measurement to be below threshold[57]. The 30-degree FOV ophthalmoscope provides a single-window from the retina, including the optic nerve head and fovea, in single-shot simultaneous MR scanning[58]. Our optical system has a manual variable focus lens and an adjustable 3D-printed ABS material holder that allows adjusting the optical system and changing the position of the lenses and distance between the device and the participant's eye to provide optimal resolution and to direct the light inside the eye without inference in positioning of the 64 channel head coils irrespective of head size and position inside the bore.

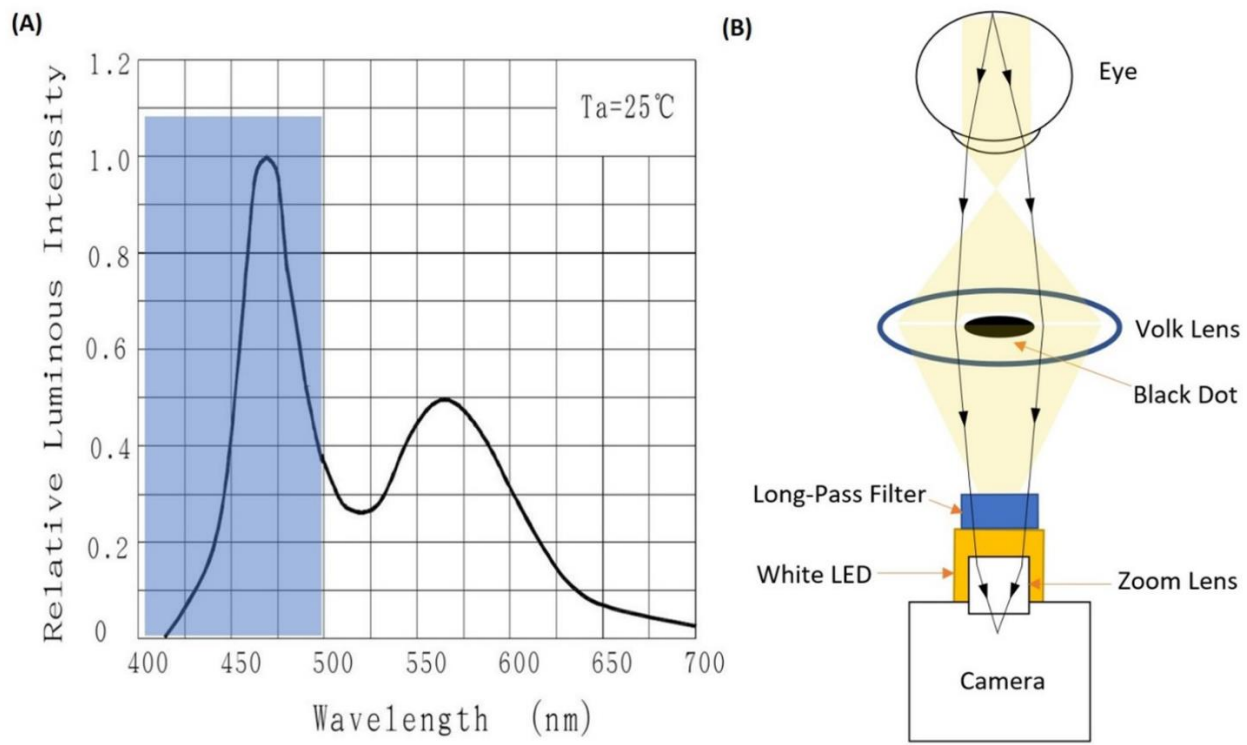

*Figure 7.* A: The light source spectrum, and the filtered region of wavelength < 500 nm. B: Optical design schematic and ray tracing.

To afford easy application of the optical device, we designed an optical setup with a single path for all light rays. Thus, the LED light projected inside the eyeball, and the light reflected from the retina follows the same path. We used a 2 mm diameter black paint dot on the surface of the Volk lens to eliminate the reflected LED light from the center of the focusing lens toward the camera, which is a problematic aspect of conventional ophthalmoscopes. The LED had the lowest light that was sustainable for long video recordings, which was directed inside the eye such that the operator could adjust the system as required during the imaging.

The maximum light intensity in the focal point area was ~7800 lux, corresponding to power of 1 mW/cm$^2$ at 555 nm wavelength. According to a previous report, the safety limit for 400 - 700 nm illumination is (a) a power of 2 mW/cm$^2$ for direct ophthalmoscopy with 500 s maximum cumulative exposure and (b) and energy of 10 mJ/cm$^2$ for times between 10 and 1000 s[57], which is an order of magnitude below damage threshold for visible laser light. On the other hand, the focal area is the entire cornea surface, which presents a surface area of about 1 cm$^2$, thus favoring the safety margin of our set-up. Also, the light source spectrum range of our white LED, there is less than ~32% energy absorption on the cornea surface, suggesting by sample calculation that illumination with 1 mW/cm$^2$ deposits only 0.32 mW/cm$^2$.

**Data pre-processing**

The data were analyzed based on the ep2d fMREye signal, MRcVO, and FEC images. The fMREye data were preprocessed and analyzed using FSL (5.09 BET software)[59], AFNI (analysis of functional neuroimages, v2)[60], and MATLAB (R2019). The functional data preprocessing was performed using the FSL pipeline. After removal of the 20 initial frames, data were high-pass filtered with a cut-off frequency of 0.008 Hz (125 s), and framewise head motion correction was then performed with FSL 5.08 MCFLIRT software[61]. The AFNI 3dDespike function was used to remove spikes from the data.

The eye camera software (MRC eye-tracking software) records video from the eye at a resolution of 640 x 480 pixel at rate of 30 frames per second, and outputs the resulting file in greyscale .vid format. We used a custom MATLAB script to extract each video frame and transform it into nifti format (.nii.gz) using the *nifti write* function. Finally, the individual frames were merged back into a video using the FSL function *fslmerge* prior to further analysis using various FSL and AFNI (Analysis of Functional Neuroimages) software.

**Data analysis**

All data from fMREye, MRcVO and FEC in nifti format were analyzed to achieve the FFT power of data using the AFNI 3dPeriodogram function. Global FFT spectra were made by calculating the mean spectrum over the whole brain using the *fslmeants* function, whereupon we used *FSLeyes* to visualize power spectra maps of eye and brain physiological pulsations. Data analysis was performed using *FSLeyes*, MATLAB, local multimodal software NAPP (nifty app for fMRI data processing) and GraphPad Prism 9 software.

The very low frequency (VLF) band for eye and brain was 0.01 - 0.1 HZ for every subject, while respiration (RESP) and cardiac (CARD) bands were 0.1 Hz wide, and centered around the individual peaks (i.e., peak ± 0.05 Hz). Source power was defined as the sum/integral over the frequency band of interest. This was calculated by extracting the frequency bins using the *fslroi* function and then summing the remaining bins with AFNI *3dTstat*. Respiratory belt data and cardiovascular SpO2 served for physiological verification of the accuracy of fMREye.

## FFT normalization procedure

As FFT spectrums of different modalities were not in the same scale, we normalized the spectra of each modality to range between 0 and 1, thus enabling visual evaluation. The values in a dataset were normalized using the following formula:

**$z_i = (x_i – min(x)) / (max(x) – min(x))$**

where, $z_i$ is the $i^{th}$ normalized value in the dataset, $x_i$ the $i^{th}$ value in the dataset, $min(x)$ is the minimum value in the dataset, and $max(x)$ the maximum value in the dataset

## Statistical analysis

The statistical significance of the slice number differences was conducted using data from the six subjects in whom we obtained 1 and 2 slice data. We used the two-tailed Student's *t*-test (significance level $p < 0.05$) for hypothesis testing between the study groups in VLF, RESP, and CARD bands (sum FFT power over the frequency band of interest) and the Kolmogorov-Smirnov test for normality of residuals. The slice thicknesses data analyses were performed using the Kruskal-Wallis one-way analysis of variance method (ANOVA) and *post-hoc* (Tukey) tests between different slice thicknesses (sum of FFT power) in VLF, RESP, and CARD bands in six subjects for the FEC synchrony fMREye data and 12 subjects for the MRcVO synchrony fMREye data. Here, we tested the normality and lognormality of the variables between groups with the

Shapiro-Wilk test. The statistical data were analyzed with MATLAB and GraphPad Prism 9 software.

## Data availability

The datasets used and/or analyzed during the current study are available from the corresponding author on reasonable request.

## Acknowledgements

Computational services from CSC and comments and corrections from Paul Cummings are cordially acknowledged.

## Author contributions

E.SM., T.J., and Ko.V. carried out scanning and data acquisitions and conceived the study. E.SM. and Ki.V. device fabrication and wrote the manuscript. H.H., E.L., and K.M. helped with the manuscript. S.V. and H.M. support scientific guidance as ophthalmologists. R.L. and H.N. provided analytical guidance. Ki.V. led the project. All authors reviewed the data and the manuscript.

## Funding

The research is funded by the **JPND2022-120 grant,** Academy of Finland's TERVA grants **314497, 335720 &** Profi-3 grant, Jane & Aatos Erkko Foundation (JAES 1& **210043**), Business Finland, and VTR-funding.

## Conflict of interest

The authors declare no competing interests.

## Additional information:

Individuals in figure 6 understand that they may be identified from photographs. The written informed consent for publication was obtained from all participants.

**Correspondence** and requests for materials should be addressed to Ki.V. or E.SM.

## Statement of Significance

We report the spectral powers of physiological eye and brain pulsation mechanisms driven by vasomotor, respiration, and cardiac rhythms in humans as measured by simultaneous fMREye and MRcVO imaging, with optimization of the fMREye parameters for that purpose. For the first time, we were able to measure the three physiological pulsations in the human eye, thus introducing a tool that should serve for future testing of eye pathologies such as glaucoma.